# Simple SIMON
## FPGA implementations of the SIMON64/128 Block Cipher

Jos Wetzels, Wouter Bokslag

`a.l.g.m.wetzels@student.utwente.nl`

`w.bokslag@student.tue.nl`

**Abstract.** In this paper we will present various hardware architecture designs for implementing the SIMON64/128 block cipher as a cryptographic component offering encryption, decryption and self-contained key-scheduling capabilities and discuss the issues and design options we encountered and the tradeoffs we made in implementing them. Finally, we will present the results of our hardware architectures' implementation performances on the Xilinx Spartan-6 FPGA series.

## 1    Introduction

In this paper we will evaluate the SIMON family of block ciphers, in particular SIMON64/128, from the perspective of hardware implementations. In section 2 we will describe the SIMON family of block ciphers in general and the block cipher SIMON64/128 in particular. In section 3 we will describe various hardware architecture designs for implementing SIMON64/128 as a cryptographic component offering encryption, decryption and self-contained key-scheduling capabilities and discuss the various design options and tradeoffs as well as the implementation issues we encountered. In section 4 we will present the performance results of our implementations on the Xilinx Spartan-6 FPGA series. Finally, in section 5 we will discuss the limitations of our work and enumerate some possibilities for future work.

## 2    The SIMON Family of Block Ciphers

SIMON is a family of lightweight balanced Feistel block ciphers designed by the NSA [1] for high-performance in hardware to address security issues for highly constrained devices. As existing cryptographic algorithms were largely designed to meet the needs of desktop computing they are not particularly suited for use in the lightweight applications that underpin the constrained platforms of so-called pervasive computing systems. The SIMON family (together with the SPECK family) have been proposed to address the growing need for flexible, lightweight cryptography.

| block size $2n$ | key size $mn$ | word size $n$ | key words $m$ | const seq | rounds $T$ |
|---|---|---|---|---|---|
| 32 | 64 | 16 | 4 | $z_0$ | 32 |
| 48 | 72 | 24 | 3 | $z_0$ | 36 |
|    | 96 |    | 4 | $z_1$ | 36 |
| 64 | 96 | 32 | 3 | $z_2$ | 42 |
|    | 128 |   | 4 | $z_3$ | 44 |
| 96 | 96 | 48 | 2 | $z_2$ | 52 |
|    | 144 |   | 3 | $z_3$ | 54 |
| 128 | 128 | 64 | 2 | $z_2$ | 68 |
|    | 192 |   | 3 | $z_3$ | 69 |
|    | 256 |   | 4 | $z_4$ | 72 |

**Fig. 1** *SIMON parameters*

The SIMON family has a number of parameters that determine its specifics as shown in figure 1. We will refer to a SIMON block cipher with an n-bit word and m-bit key as SIMON2n/m, ie. a configuration with 32-bit words and a 128-bit key becomes SIMON64/128.

### 2.1 Configuration

We chose SIMON64/128 as our cipher configuration for the architecture implementations in our paper. While block sizes smaller than 128 bits can offer sub-standard security in certain scenarios (eg. short cycle problems in OFB mode, distinguishing attacks, etc.) we aimed at our implementation being compatible with the Xilinx Spartan-6 [2] family of FPGAs and as such were restricted with regards to the number of available IOBs/pins. Our designs, however, are easily extendible to the SIMON128/128 configuration offering a fully adequate security level. In the rest of this paper, our SIMON parameters will be as follows:

- *Block size*: 64
- *Key size*: 128
- *Word size*: 32
- *Key words*: 4
- *Constant sequence*: $z_3$
- *Rounds*: 44

## 2.2 Round Function

The SIMON round function is an AND-RX construction with a balanced Feistel structure that utilizes the following operations:

- Bitwise XOR, denoted as $x \oplus y$.
- Bitwise AND, denoted as $x \& y$.
- Left bitwise rotation ROL, denoted as $S^y(x)$ where $y$ is the rotation count.

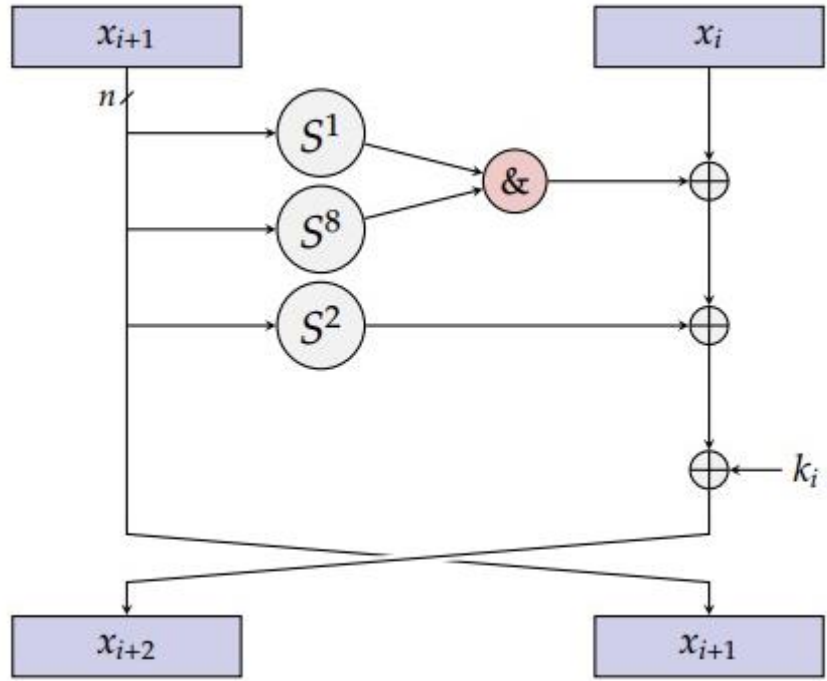

**Fig. 2** *SIMON round function*

The SIMON round function (used for encryption) can be expressed as:

$$R(l, r, k) = ((S^1(l) \& S^8(l)) \oplus S^2(l) \oplus r \oplus k, l)$$

And it's inverse (used for decryption) as:

$$R^{-1}(l, r, k) = (r, (S^1(r) \& S^8(r)) \oplus S^2(r) \oplus l \oplus k)$$

Where $l$ is the left-most word of a given block, $r$ the right-most word and $k$ the appropriate round round key.

## 2.3 Key Schedule

The SIMON key schedule provides key expansion capabilities by subsequently generating all round keys from the master key. In our chosen SIMON64/128 configuration the key schedule generates 44 32-bit sized round keys from the initial 128-bit master key. It does so by, for a given round $i$, combining the currently cached previous $n$ round keys (where $n$ is the key words parameter) with constant $c$ and a 1-bit round constant. The key expansion function utilizes the following operations:

- Bitwise XOR, denoted as $x \oplus y$.
- Right bitwise rotation ROR, denoted as $S^{-y}(x)$ where $y$ is the rotation count.

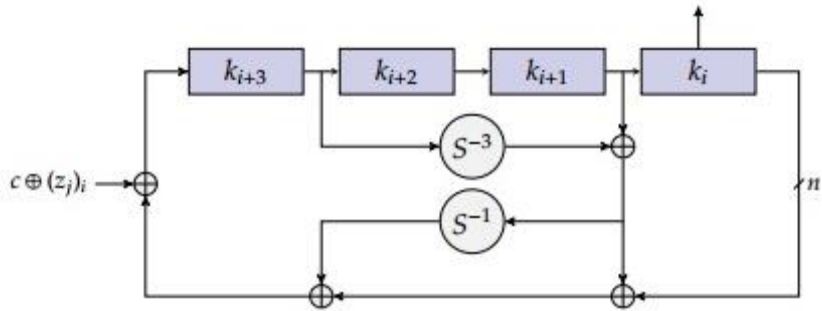

**Fig. 3** *SIMON 4-word key expansion*

The key expansion function can be expressed as:

$$K_i(k, c, z_j) = F(k_{i+3}, k_{i+1}) \oplus S^{-1}(F(k_{i+3}, k_{i+1})) \oplus k_i \oplus c \oplus (z_j)_i$$

Where:

$$F(x, y) = S^{-3}(x) \oplus y$$

And where after each expansion operation, the cached round keys are rotated rightward with the first being discarded and the last being replaced by the newly generated round key, ie.: $k_i = k_{i+j}$ for $j = 0,...,2$ and $k_{i+3} = K_i(k, c, z_j)$.

The SIMON key schedule employs a sequence of 1-bit round constants $z_i$ for purposes of the elimination of slides properties and circular shift symmetries. In our configuration that is (represented in little-endian):

$z_3$ =11110000101100111001010001001000000111101001100011010111011011

In addition, the key schedule employs the constant $c = 2^n - 4 = $ 0xFF..FC where n is the word size parameter, hence $c = 2^{32} - 4$ in our configuration.

## 2.4 Encryption

Encryption of a 64-bit plaintext block $p$ simply consists of 44 applications of the round function with the respective round key produced by the key schedule. Due to the nature of the round and key expansion functions they can be run in parallel if so desired.

## 2.5 Decryption

Decryption of a 64-bit ciphertext block $c$ consists of first swapping the left- and right-most 32-bit words followed by 44 applications of the round function but with round keys in reverse order (ie. round keys 43,...,0) followed by a final swapping of the left- and right-most words.

# 3 Hardware Design

In this section we will discuss the various hardware architectures in which SIMON can be implemented and the associated design options, issues and tradeoffs. Our implementations were designed with Field Programmable Gate Array (FPGA) usage in mind, particularly the *Xilinx Spartan-6* family. FPGAs consist of a multitude of (re)configurable universal slices which are connected in (re)configurable ways. The reconfigurable nature of the FPGA allows designers to implement various totally different functions on the same device. All discussed hardware architectures were implemented by us in VHDL [3], simulated using *Mentor Graphics ModelSIM PE* and performance- and synthesis-tested using *Xilinx ISE Design Suite 14.7*.

Our implementations were not optimized with regards to particular performance characteristics but rather serve to illustrate the options offered and problems posed by various hardware architectures as applied to SIMON and as a general indication of the cipher's performance. As opposed to other work on the implementation of the SIMON family [4,5,6,7] we decided to implement each architecture as a fully functional cryptographic component offering both encryption and decryption as well as self-contained key-scheduling capabilities.

## 3.1 Tradeoffs

In this sub-section we'll briefly investigate several design tradeoffs we considered when implementing the architectures discussed further on as well as some tradeoffs to be considered by those seeking to build their own implementation of the given architectures.

### 3.1.1 Dimensions of Parallelism

As discussed by Aysu et al. [5] the block cipher design space offers several dimensions of parallelism: *rounds*, *encryptions* and *bits*. The particular parallelism choices affect the performance results (both area and throughput) of a given cipher implementation.

- *Parallelism of Rounds:* Within a given encryption component the number of rounds executed in parallel $r$ can range from $r = 1$ to $r = \#rounds$. Increasing round parallelism requires corresponding (partial to full) loop unrolling and if so desired outer-round pipelining as discussed in section 3.6. Given that an increase in round parallelism comes with an increase in area, the choice for the degree of round parallelism $r$ depends on a throughput/cost tradeoff that can be determined from the throughput-to-area ratio. We chose to implement full loop unrolling and outer-round pipelining as area limit didn't play a role in our implementations.

- *Parallelism of Encryptions:* Given enough available area on the target FPGA one can use $e$ separate encryption components in parallel thus linearly increasing the overall system throughput. When throughput maximization, rather than cost/area minimization, is the primary concern and the target platform limit allows for it encryption parallelism of a suitable architecture with the best throughput-to-area ratio is recommended.

- *Parallelism of Bits:* Within a given round operation the input size $i$ of the operators that make up its combinational logic can range from 1 to $n$ bits (where $n$ is the cipher block size). A round implemented with $i = 1$ is called *bit-serialized* while a round with full parallelism (ie. $i = n$) is called *iterated* and processes a full block during every clock cycle. Obviously $i$ is positively related to both throughput and area. Given the design specifications of SIMON64/128 and the capabilities offered by the Xilinx Spartan-6 FPGA family we chose to implement all our designs as *iterated* designs with respect to bit-parallelism.

### 3.1.2 Block Cipher Modes of Operation

Block ciphers are used in a so-called mode of operation which combines individual ciphertext blocks derived from individual plaintext blocks into a single ciphertext. As noted by Gaj et al. [8] certain block cipher design architectures lend themselves better to usage with certain modes of operation than others. Modes of operation can be divided into two main categories:

- *Non-feedback modes*: Where encryption of subsequent plaintext blocks can be performed independently from other blocks (eg. ECB, CTR, etc.)

- *Feedback modes*: Where encryption of a block cannot start until encryption of the previous block is finished (eg. CBC, CFB, OFB, etc.)

Hence feedback modes of operation require sequential processing without allowing for parallelism. As such those wishing to utilize a SIMON64/128 implementation for use in feedback modes could choose the loop unrolling architecture but should avoid the pipelining architectures.

### 3.2 Round Function

We first implemented the SIMON round function as a standalone component and reference for subsequent architectures. We implemented it as a combinational circuit directly expressing the round function $R$ as defined in section 2.2.

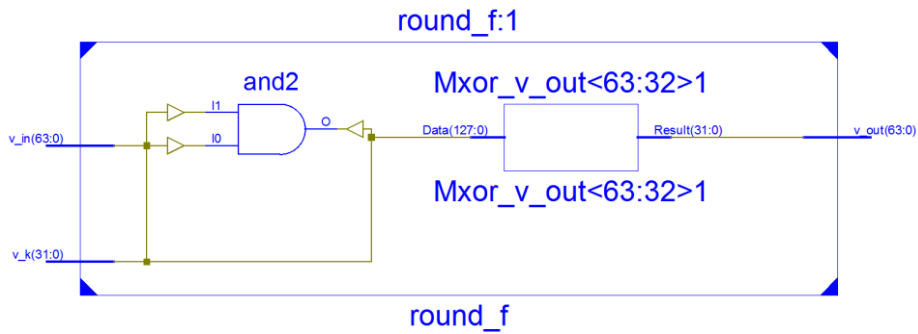

**Fig. 4** SIMON64/128 *round function as combinational circuit*

### 3.3 Iterative Architecture

We implemented SIMON64/128 as a basic iterative architecture. In the iterative architecture [9] the round function is implemented as a combinational circuit joined with a single register and multiplexer and connected to a signal feeding it the appropriate round key.

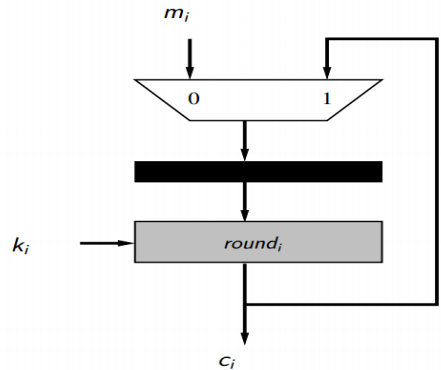

**Fig. 5** *Basic iterative architecture (courtesy of slides by A. de la Piedra [10])*

During the first clock cycle the plaintext block is fed to the circuit and stored in the register and with each subsequent clock cycle a cipher round is executed and its result is fed back into the circuit through the register. After $n$ clockcycles (where $n$ is the number of rounds) the register now holds the ciphertext block corresponding to the plaintext and key combination. As such only a single block of data is encrypted at a time and encryption of a plaintext block takes a number of clock cycles equal to the number of cipher rounds.

### 3.3.1 Encryption and Decryption

Our iterative design supports both encryption and decryption functionality which can be selected with a single-bit signal. Due to the balanced Feistel nature of SIMON, encryption and decryption are symmetric (save for the reverse order of round keys) and as such no additional components or circuitry is required for decryption functionality. The reversed round key scheduling, however, does pose a design problem.

Given that the key schedule is iterative and every round key is derived from the previous 4 round keys we can run encryption and key expansion parallel, feeding the correct round key to the round function each cycle. However, since decryption consumes the final round key as the first this means that we need to have pre-expanded all round keys before decryption commences since in order to generate the final round key we need to have already generated all others. In order to address this we decided to add a RAM component to our iterative design. The RAM component holds 44 32-bit sized word cells to store the round keys which can then be written to or read from RAM as required.

We have roughly two design options to integrate the RAM into our iterative architecture:

a) *Separate*: Pre-expansion is a *separate* phase next to the regular *initialization* and *run* phases where the SIMON component will run for $n = 44$ clock cycles each of which generates the corresponding round key and stores it in RAM. Subsequent encryption or decryption functionality will read the appropriate round key from RAM based on the round index. This approach introduces a slight area penalty as well as requiring both encryption and decryption to be prefaced with $n$ additional pre-expansion cycles.

b) *Integrated*: Pre-expansion is *integrated* into encryption functionality since during encryption key expansion can run in parallel and generated round keys can be stored in RAM as they are generated. This means that no separate pre-expansion phase is required for encryption and that decryption can simply be prefaced with 44 additional rounds of encryption over a block of bogus data to pre-expand the key in memory for subsequent consumption by decryption rounds.

We refer to the evaluation in section 4 for performance details.

### 3.3.2 Key Schedule

We chose to implement the key schedule as a combinational circuit generating round keys on the basis of a supplied round index $r$ and the left-shifting cache of the previous 4 round keys $k_i, .., k_{i+3}$. We also chose to conflate the $c$ and $z_3$ constants to a single constant $C = c \oplus z_3$ for efficiency purposes.

Given our requirement for round key storage in RAM, there are two different models of connecting key scheduling to the round function:

a) *RAM-routing*: We connect the round keys output by the key schedule to the input of the RAM module and connect the output of the RAM module to the round function, introducing a single clock cycle delay between round key generation and consumption. In order to address this the initialization phase takes 2 clock cycles to align round keys with rounds and the key schedule will generate all round keys $k_i$ ($i \in \{0,...,43\}$). In addition, the key schedule output is also connected to the final word of the round key cache.

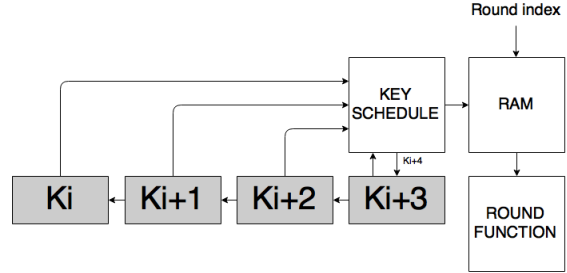

**Fig. 6** *Ram-routing approach*

b) *Cache-routing*: We connect the first word of the round key cache to the round function and connect the output of the key schedule to the last word of the round key cache. In this fashion, round key $k_{i+4}$ will be fed into the left-shifting cache in time for proper subsequent round key generation. The first word of the round key cache is also connected to the input of the RAM module to fill it with the pre-expanded key for subsequent decryption operations. The first word of the round key cache is connected via a multiplexer to both the second word of the cache (for operating in encryption mode) and the output of the RAM module (for operating in decryption mode). In this model the initialization phase takes only a single clock cycle but if the encryption mode is used for key pre-expansion purposes for subsequent decryption an additional cycle is needed to completely fill the RAM. In the cache-routing approach the key schedule will generate round keys $k_i$ ($i \in \{4,..,43\}$).

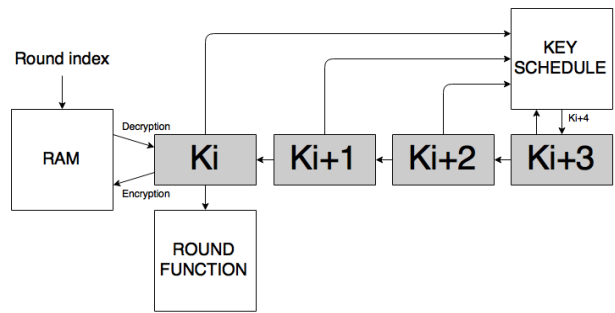

**Fig. 7** *Cache-routing approach*

We implemented the *integrated* pre-expansion mentioned in section 3.3.1 with both models (*RAM-routing* and *cache-routing*) and implemented the *separate* pre-expansion method with *RAM-routing*. We refer to the evaluation in section 4 for performance details.

### 3.4 Loop Unrolling Architecture

We implemented an instance of SIMON64/128 as a (full) loop unrolling architecture. In the loop unrolling architecture [9] single combinational parts of the circuit of an iterative architecture are '*unrolled*' to implement $K$ rounds (where $1 \leq K \leq$ #$rounds$ and $K$|#rounds) of the cipher instead of a single round (with key scheduling being unrolled in similar fashion). Hence the number of clock cycles necessary to encrypt or decrypt a block of data is decreased by a factor of $K$ and the minimum clock period is decreased by a factor slightly less than $K$ giving an overall increase in throughput and decrease in latency while simultaneously resulting in an increase in area more or less proportional to $K$ due to unrolling of combinational logic of round and key expansion functionality as well as the number of simultaneously stored round keys.

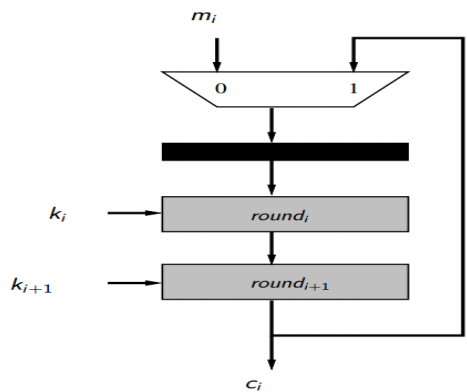

**Fig. 8** *Loop unrolling architecture for $K = 2$ (courtesy of slides by A. de la Piedra [10])*

In loop unrolling one has the choice between *partial* ($K <$ #$rounds$) and *full* ($K =$ #$rounds$) unrolling where one has to make a tradeoff between throughput and area. Given that our Spartan-6 target platform is well-equipped to handle the maximum area increase that comes with full unrolling we chose to implement full unrolling in order to achieve maximum throughput for this particular architectural mode. It is of course possible to scale back our design to a partial unrolling architecture if so desired by re-introducing the feedback loop and multiplexer of the basic iterative architecture.

As noted by Gaj et al. [9], however, full loop unrolling is recommended only for block ciphers operating in feedback modes of operation in implementations which can tolerate large circuit area increases.

### 3.4.1 Round Function

We replicated the round $K = 44$ times to implement full unrolling, inter-connecting each round with the next and connecting every round to the appropriate signal delivering the round key from the unrolled key scheduling circuit. In this manner the plaintext is transformed into ciphertext by executing 44 round function operations in a single clock cycle.

### 3.4.2 Key Schedule

Key scheduling was fully unrolled by replicating the scheduling function $K = 44$ times and both connecting the round key carrying output of every unrolled operation to the appropriate unrolled round and inter-connecting the unrolled key scheduling operations in order to make sure all round keys are generated in a single clock cycle.

### 3.4.3 Encryption and Decryption

Encryption and decryption functionality differs from the iterative architecture in that no RAM is required anymore (due to full unrolling of the key schedule). Hence in the full loop unrolling architecture, decryption does not require a pre-expansion step. In addition, both encryption and decryption require only a single clock cycle.

## 3.5 Inner-Round Pipelining Architecture

We implemented an instance of SIMON64/128 as an inner-round pipelining architecture derived from the iterative architecture (of the *integrated, cache-routing* variety) described in section 3.3. In the inner-round pipelining architecture [9] the designer starts out with the basic iterative architecture and performs the following steps:

- The round function is divided into $n$ independent sub-functions.
- $K$ registers are inserted between the round sub-functions where $1 \leq K \leq n$.
- The optimal $K$ is determined that balances throughput and area.

The insertion of registers inside a cipher round increases throughput while only minimally increasing area, resulting in an overall increase of the throughput-to-area ratio up until the optimal value for $K$ (after which throughput may keep increasing but the throughput-to-area ratio will start decreasing). In the inner-round pipelining architecture the designer has to find the optimal $K$ (within area constraint bounds) that achieves the best throughput-to-area ratio.

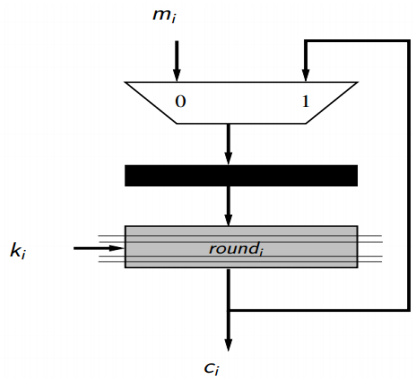

**Fig. 9** *Inner-round pipelining architecture for $K = 4$ (courtesy of slides by A. de la Piedra [10])*

During our design and implementation of inner-round pipelining of SIMON64/128 we encountered several limits which we will discuss in section 3.5.1.

### 3.5.1 Round Function

We started out by constructing a partitioning tree of the SIMON round function $R(l, r, k) = ((S^1(l) \: \& \: S^8(l)) \oplus S^2(l) \oplus r \oplus k, l)$:

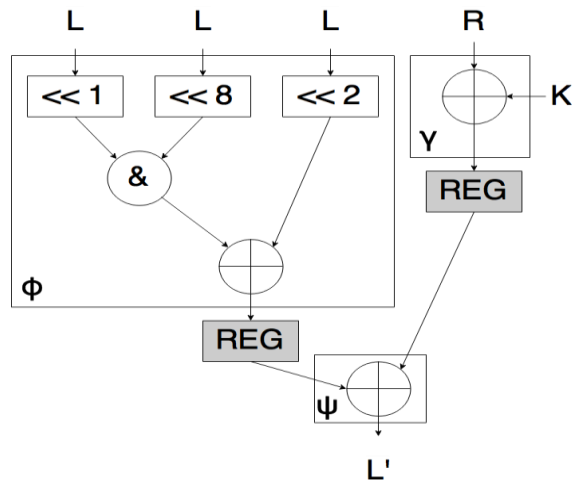

**Fig. 10** *SIMON round function partitioning showing options for inner-round register insertion.*

As shown by figure 10, the SIMON round function can be divided into 3 independent sub-functions:

$$\gamma(r,k) = r \oplus k$$
$$\varphi(l) = (S^1(l) \mathbin{\&} S^8(l)) \oplus S^2(l)$$
$$\psi(x,y) = x \oplus y$$

So that:

$$R(l,r,k) = (\psi(\varphi(l), \gamma(r,k)), l)$$

Obviously, semantically $\gamma = \psi$ and as such requires only a single component definition in the VHDL specification. This division offers 3 (since there already is a register supplied by the iterative architecture before $\gamma$ and after $\psi$) potential options for inner-round register insertion, namely: after $\gamma$, after $\varphi$ and after both.

Since the SIMON family of block ciphers are Feistel networks this limits our options for inner-round pipelining due to the fact that at the beginning of a round the right word is to be the previous left word and the left word the result of the previous right word put through the round function. Since inner-round pipelining delays the availability of the round function output by $K$ clock cycles this means that the left word (derived from the round function) for a given round cannot be available in time which prohibits pipelining round sub-functions derived from it (such as only inserting a register after $\varphi$ without a balancing register after $\gamma$). As such we did not further investigate options for pipelining the operations internal to the sub-function $\varphi$.

We explored options for using 32-bit negative-edge triggered registers (so that during every clock cycle starting at a rising edge the register consumes the output of the pipelined sub-function after the falling edge and it is available at the end of the same cycle) but due to the deadline constraints of this project and the overall absence of detailed literature on the matter of inner-round pipelining for Feistel networks, we have restricted ourselves to implementing experimental inner-round pipelining designs for the options $K = 1$ (register insertion after $\gamma$) and $K = 2$ (register insertion after both $\varphi$ and $\gamma$). We did not, however, have time to implement test benches or thoroughly verify their correctness and as such have not included corresponding performance figures in section 4.

### 3.6  Outer-Round Pipelining Architecture

We implemented an instance of SIMON64/128 as an outer-round pipelining architecture derived from the loop unrolling architecture described in section 3.4. In the outer-round pipelining architecture [9] the designer determines the number of rounds $K$ that can be loop-unrolled without exceeding the maximum designated (either through availability or other restrictions) circuit area and inserts registers inside the combinational circuit between all subsequent cipher rounds.

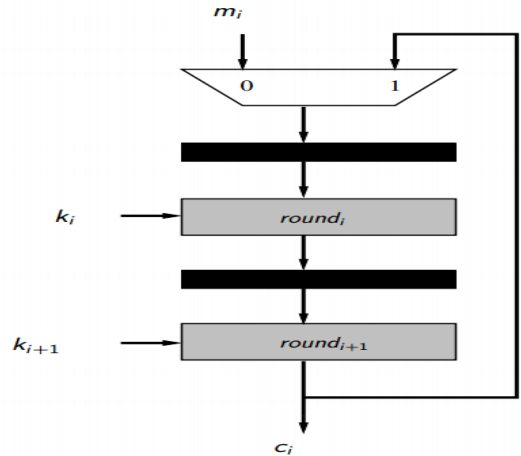

**Fig. 11** *Outer-round pipelining architecture for $K = 2$*
*(courtesy of slides by A. de la Piedra [10])*

This allows for the simultaneous processing of $K$ blocks of data by the outer-round pipelined circuit (given that each of the $K$ blocks can be stored at the $K$ different registers at the end of a given clock cycle) leading to a corresponding increase in both throughput and area. Given that we established in section 3.4 that full loop unrolling is possible within our area constraints we decided to implement full outer-round pipelining.

#### 3.6.1 Encryption and Decryption

Round, key expansion, encryption and decryption functionality was implemented identical to the full loop unrolling architecture with the addition of a series of 43 (due to the final round being connected directly to the output as opposed to a feedback register) 64-bit positive-edge triggered registers in between the unrolled rounds in the encryption and decryption component. While encryption and decryption both take $\#rounds$ clock cycles to complete, as in the iterative architecture, in the latter new blocks of plaintext can be fed into the system only once every $\#rounds$ clock cycles while in the full outer-round pipelining architecture a new plaintext block can be fed into the system every clock cycle.

### 3.7 Mixed Inner-Outer-Round Pipelining Architecture

We implemented an instance of SIMON64/128 as a mixed inner-outer-round pipelining architecture derived from the outer-round pipelining architecture described in section 3.6. Given that mixed pipelining offers a significant improvement on throughput compared to outer-round pipelining and the circuit area limit is large enough we can use our fully unrolled outer-round pipelining architecture as the based for the mixed architecture.

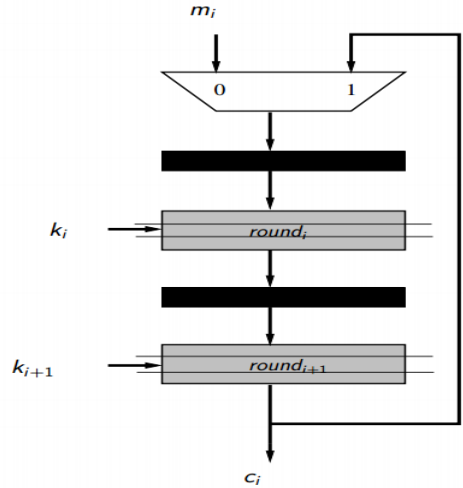

**Fig. 12** *Mixed pipelining architecture for $K_i = 2, K_o = 2$*
*(courtesy of slides by A. de la Piedra [10])*

In the mixed pipelining architecture [9] the designer starts out with a partially or fully unrolled outer-round pipelining architecture and replaces the round function by the round function of the optimal case $K$ found in the inner-round pipelining architecture, giving an architecture with $K_i$ inner-round registers and $K_o$ outer-round registers. Given that our inner-round pipelining designs in section 3.5 were, as of publication, merely experimental and still untested we established a provisional optimum of $K = 2$ and implemented a mixed pipelining architecture with $K_i = 2, K_o = 43$. We developed test-benches that confirm the correctness of our implementation but due to the untested nature of the inner-round pipelining design the performance results obtained by the mixed pipelining architecture implementation are to be considered purely provisional.

### 3.7.1  Encryption and Decryption

The round function was implemented identical to that of the inner-round pipelining architecture for $K = 2$ while key expansion, encryption and decryption functionality where implemented identical to the outer-round pipelining architecture. Encryption and decryption both take 1 additional clock cycle compared with the outer-round pipelining architecture and a new plaintext block can be fed into to the system every clock cycle.

## 4  Performance

In this section we provide an overview of the performance results of the implementations discussed in section 3. The area and throughput performance figures were derived either directly or partially from the results obtained by full design implementation of our VHDL code using *Xilinx ISE Design Suite 14.7*.

We have refrained from a comparison between the performance results of our implementations and those of other SIMON implementations or similar lightweight block ciphers. This is primarily because this work does not seek to present particularly optimized implementations but also since (as noted elsewhere [8]) it is inherently difficult to perform such a comparison reliably and meaningfully since different authors implement their designs under different assumptions and with different optimization goals (eg. different platform assumptions, different levels of parallelism, different optimization goals, unclarity regarding implementation completeness: is decryption and key scheduling implemented or not, etc.). As such the results presented in this section serve as a guide to choosing the appropriate design architecture for implementing SIMON.

### 4.1  Area

The area required by cipher implementations is an important parameter since it is positively correlated with production cost and the viability of implementation on a certain platform is determined by its area limit.

| Architecture | Area (#occupied slices) | Area (#slice registers) | Area (#slice LUTs) |
|---|---|---|---|
| Round Function | 24 | 0 | 32 |
| Iterative* | 148 | 201 | 353 |
| Iterative** | 105 | 202 | 329 |
| Iterative*** | 113 | 199 | 328 |
| Full loop unrolling ($K = 44$) | 2952 | 0 | 8096 |
| Full Outer-Round Pipelining ($K = 44$) | 1149 | 2752 | 4096 |
| Mixed Pipelining ($K_i = 2, K_o = 44$) | 2161† | 5568† | 6459† |

**Table 1** *Area results*
\* = (*Integrated pre-expansion, Cache-routing*), \*\* = (*Integrated pre-expansion, Ram-routing*), \*\*\* = (*Separate pre-expansion, Ram-routing*) , † = provisional results.

Since our designs were implemented with FPGA usage in mind we will measure the required area in so-called *occupied slices* which are the number of basic FPGA blocks occupied by our implementation. A slice contains a given number of LUTs (set of hardwired logic gates), flip-flops, multiplexers, etc. and is differently implemented depending on the FPGA family. Table 1 lists the total number of occupied slices (and the number of slice registers and slice LUTs that they are composed of as reported by *Xilinx ISE*) for each design implementation.

### 4.2 Throughput

Throughput measures the number of bits that can be processed (eg. encrypted or decrypted) by the component in a certain unit of time. Given the Feistel network nature of the SIMON family and hence the equality of encryption and decryption round functions, throughput rates for encryption and decryption are equal and hence we will report throughput as a single parameter. Table 2 lists the various architectures, the formula for determining their throughput and the actual throughput rates. Keep in mind that throughput rates for the round function concern a single round not a full block encryption.

| Architecture | Throughput | Throughput rate (Mbit/s) |
|---|---|---|
| Round Function | $\frac{blocksize}{T_{clk}}$ | 9985.957 |
| Iterative* | $\frac{blocksize}{\#rounds * T_{clk}}$ | 242.303 |
| Iterative** | $\frac{blocksize}{\#rounds * T_{clk}}$ | 206.524 |
| Iterative*** | $\frac{blocksize}{\#rounds * T_{clk}}$ | 269.760 |
| Full loop unrolling ($K = 44$) | $\frac{blocksize}{\frac{\#rounds}{K} * T_{clk}}$ | 479.300 |
| Full Outer-Round Pipe-lining ($K = 44$) | $\frac{K * blocksize}{\#rounds * T_{clk}}$ | 33109.157 |
| Mixed Pipelining ($K_i = 2, K_o = 44$) | $\frac{K_o * blocksize}{\#rounds * T_{clk}(K_i)}$ | 11666.059† |

**Table 2** *Throughput results*
\* = (*Integrated pre-expansion, Cache-routing*), \*\* = (*Integrated pre-expansion, Ram-routing*), \*\*\* = (*Separate pre-expansion, Ram-routing*), † = provisional results, $T_{clk}$ = *minimum clock delay*.

### 4.3 Throughput-To-Area Ratio

Different applications require different optimization goals. For example, for many lightweight applications (eg. cryptography in RFID applications) throughput itself is not the top priority as only relatively small volumes of data are to be processed. Given that in general, however, we seek to maximize throughput and minimize area but that both are positively related (ie. an increase in throughput usually requires a corresponding increase in area), one usually has to look at the throughput-to-area ratio (expressed in the number of bits processed per time unit per area unit) rather than either parameter in isolation. If we combine the performance results in table 1 with those in table 2 we get throughput-to-area ratio table 3. Keep in mind that throughput rates for the round function concern a single round not a full block encryption.

| Architecture | Throughput rate (Mbit/s) | Area (#occupied slices) | Throughput-To-Area (Mbit/s per slice) |
|---|---|---|---|
| Round Function | 9985.957 | 24 | 416.082 |
| Iterative* | 242.303 | 148 | 1.637 |
| Iterative** | 206.524 | 105 | 1.967 |
| Iterative*** | 269.760 | 113 | 2.387 |
| Full loop unrolling ($K = 44$) | 479.300 | 2952 | 0.163 |
| Full Outer-Round Pipelining ($K = 44$) | 33109.157 | 1149 | 28.816 |
| Mixed Pipelining ($K_i = 2, K_o = 44$) | 11666.059† | 2161† | 5.398† |

**Table 3** *Area results*

\* = (*Integrated pre-expansion, Cache-routing*), \*\* = (*Integrated pre-expansion, Ram-routing*), \*\*\* = (*Separate pre-expansion, Ram-routing*), † = provisional results.

## 5 Conclusion and Future Work

In this paper we presented several architectural designs for implementing the SIMON64/128 block cipher and the options and tradeoffs that come with these designs and the issues encountered during their implementation. We presented their performance results which show the SIMON family of block ciphers is well-suited as a candidate for applications requiring flexible, lightweight cryptography.

As discussed in sections 3.5.1 and 3.7 due to project deadline constraints we were unable to properly validate the implementations of our inner-round pipelining designs and their corresponding performance figures. As a result both the implementation and performance figures of the mixed pipelining architecture are to be considered provisional. The finalization of these implementations, their thorough testing and the inclu-sion of their performance figures for comparison with those presented in section 4 are left to future work.

Neither our designs presented in section 3 nor their specific implementations [3] and the derived performance results in section 4 were optimized with respect to any performance metric or parameter. As such the performance results could be improved by introducing such optimizations as has been done in related work [5].

Our designs have not been hardened against possible side-channel attacks (eg. power analysis, fault injection, etc.) or tampering efforts. Given the threat posed by such attacks, research in this direction (as done by related work [7]) is a fruitful area for future work.